# Readout Electronics for CBM-TOF Super Module Quality Evaluation

Wei. Jiang, Xiru Huang, Ping Cao, Chao Li, Junru Wang, Jiawen Li, Jianhui Yuan, Qi An

***Abstract*—A super module assembled with MRPC detectors is a component of TOF (Time of Flight) system for the Compressed Baryonic Matter (CBM) experiment. Before the super modules are applied to CBM-TOF, their quality needs to be evaluated. The readout electronics is confronted with a tremendous challenge of transmitting data at the maximal speed of 6 Gbps. In this paper, the readout method for CBM-TOF super module quality evaluation is presented. A parallel architecture based on the Gigabit Ethernet is designed to meet the requirement for data transmission rate. First, the data is sent from the front-end electronics to four readout module groups via optical fibers at the maximal rate of 1.5 Gbps per fiber. Next, the data are further distributed to sixteen parallel daughter readout models so that the maximal data throughput of each daughter readout module is 375 Mbps, within its capacity of 550 Mbps. Finally, the readout daughter modules send data to the data acquisition (DAQ) software through standard Gigabit Ethernet. The proposed readout method has the advantage of good scalability, so it can meet different requirements for variable data rate. The preliminary test result shows that each of four parallel readout groups can transmit data at the speed of 1.6 Gbps, which indicates that the overall readout system can meet the requirement for the maximal data-transmission speed of 6 Gbps.**

## I. Introduction

THE COMPRESSED Baryonic Matter (CBM) experiment aims at exploring the phase diagram of strongly interacting matter [1] [2]. TOF (Time of Flight) system is one of the key parts in hadron identification. The TOF system has a wall structure shown in Fig. 1. Six types of super modules labeled as M1-M6 which are assembled with multiple Resistive Plate Chambers (MRPCs) are the component unit of CBM-TOF system [3][4].

Before these super modules are applied to the CBM-TOF, their quality has to be tested by the evaluation electronics. In M5 and M6, there are 5 MRPCs, each of which has 64 electrical readout channels. Thus the total number of electrical readout channels in both M5 and M6 modules is 320 channels. According to the Monte Carlo simulation result, the maximal event rate per channel of M5 and M6 is 300 kHz and 200 kHz, respectively [5]. The time resolution of the Time-to-Digital Converter (TDC) is expected to be less than 20 ps (RMS) and 32 ps (bin size) [5]. Thus, more data bits are required to encode high -precision TOF measurement results. The width of the TDC measurement result is 48 bits in the readout electronics [6].

Manuscript received June 23, 2018; revised January 09, 2019; accepted February 11, 2019.

This work was supported by the National Basic Research Program (973 Program) of China under Grant 2015CB856906.

The authors are with the State Key Laboratory of Particle Detection and Electronics, University of Science and Technology of China, Hefei 230026, China (email: xiru@ustc.edu.cn).

W. Jiang, J. Li and J. Yuan are with the Department of Engineering and Applied Physics, University of Science and Technology of China, Hefei 230026, China.

P. Cao, X. Huang, C. Li, J. Wang and Q. An are with the Department of Modern Physics, University of Science and Technology of China, Hefei 230026, China.

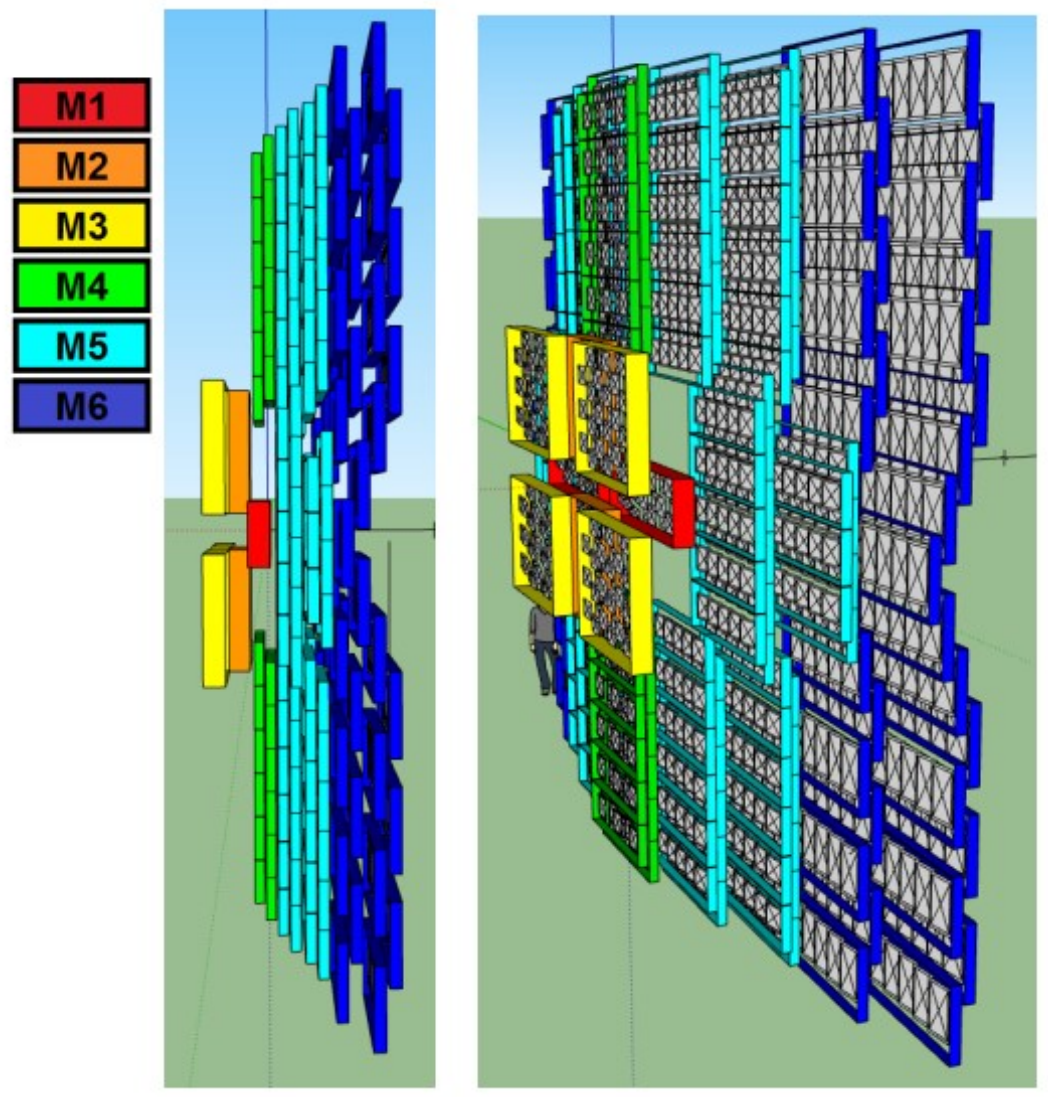

Fig. 1. Side view (left) and oblique rear view (middle) of the TOF wall.

Based on the above analysis, the maximal raw data rate of a super module M5 can be calculated as follows:

$$300\ \text{kHz/ch} \times 320\ \text{ch} \times 48\ \text{bit} = 4.6\ \text{Gbps}. \quad (1)$$

Taking the cost of transmission protocol encapsulation into consideration, the actual maximal data rate of a super module is about 6 Gbps, which puts a enormous pressure on the readout electronics.

Generally, in the particle experiments such as BESIII [7], STAR [8], CSNS-WNS [9], etc., the data from the front-end electronics are first aggregated at several readout modules which are assembled on the Peripheral Component Interconnect extensions for Instrumentation (PXI) or Versa Module Eurocard (VME) crates, and then transmitted to the crate controller through the crate backplane. Finally, the crate controller sends the data to the data acquisition (DAQ) software. However, due to the limitation of the crate backplane bandwidth, the traditional VME or PXI technique cannot satisfy the requirement for the maximal data reading speed of 6 Gbps of the CBM-TOF experiment.

In this paper, a new readout method for quality evaluation of the CBM-TOF super modules is proposed. In the proposed method, data from front-end electronics are transmitted to the DAQ software through the Gigabit Ethernet ports on the distributed readout modules, instead of through the crate backplane or crate controller. Furthermore, a parallel architecture is designed to share the transmission demand equally among all the readout modules. Moreover, the proposed method has good scalability when the data rate increases, and is not limited by the bandwidth of PXI or VME crate backplane.

## II. Architecture Of Readout Electronics

### *A. Front-end electronics*

In every M5 and M6 super module, there are 10 PADI (Pre-Amplifier and DIscriminator) boards and 5 MRPCs [10], which constitute a total of 320 readout channels. The MRPCs generate weak analog signals which are converted to a low-voltage differential signal (LVDS) after discrimination by the PADI boards. The PADI is designed according to the time over threshold (TOT) technique, which means that signal leading edge encodes the arriving time of a particle, and signal width encodes the particle's charge saved in the MRPCs.

In [6], it is shown that the front-end electronics named the sandwich TDC station (STS) serves for receiving 320-channel TOT signals and sending the digitized time data to the back-end electronics.

The architecture of the STS [6] is shown in Fig. 2. Every STS comprises a TOT feeding board (TFB), ten TDC cards, and a TDC readout motherboard (TRM). One side of the TDC card is connected to the TFB, and the other side is connected to the TRM via gold fingers. The TFB transfers the 320-channel TOT signals from a super module equally distributing them to ten TDC cards. The TRM receives data from ten TDC cards and sends it to the back-end readout electronics through four optical fibers with small form-factor pluggable (SFP) transceivers.

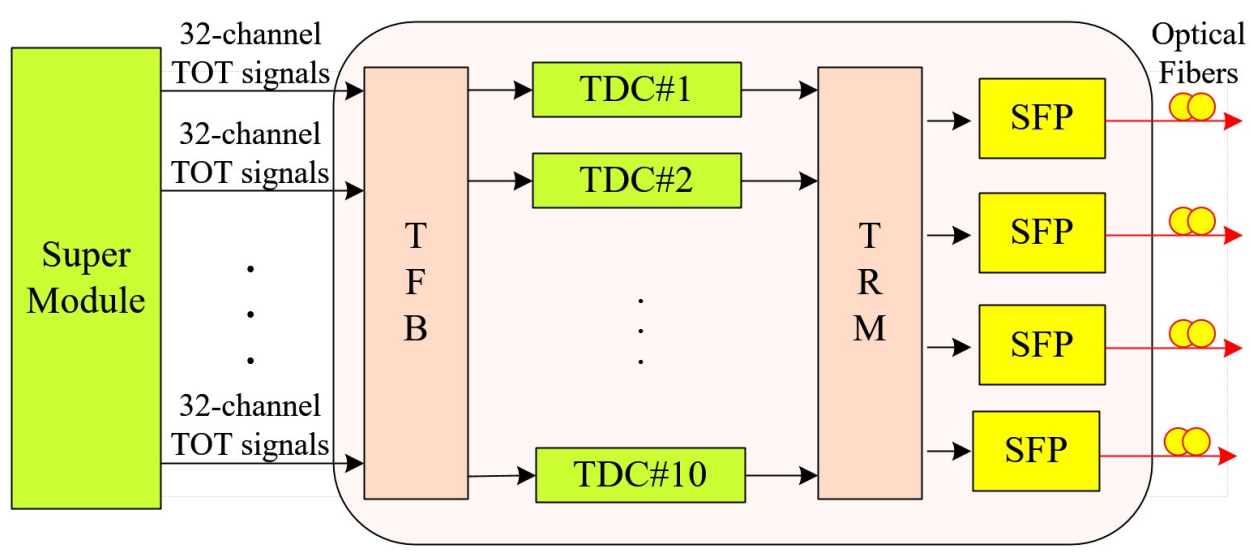

Fig. 2. The STS architecture.

### *B. Back-end electronics*

As can be seen in Fig. 3, there are two types of readout modules in the back-end electronics: DRM (Data Readout Module) and DMB (DRM Mother Board). The DRM is used to support the Gigabit Ethernet transmission, and it sends data directly to the DAQ software through the TCP/IP Gigabit Ethernet. The DMBs, having two DRMs, are designed to re-distribute data to the DRMs to share the required data transmission rate.

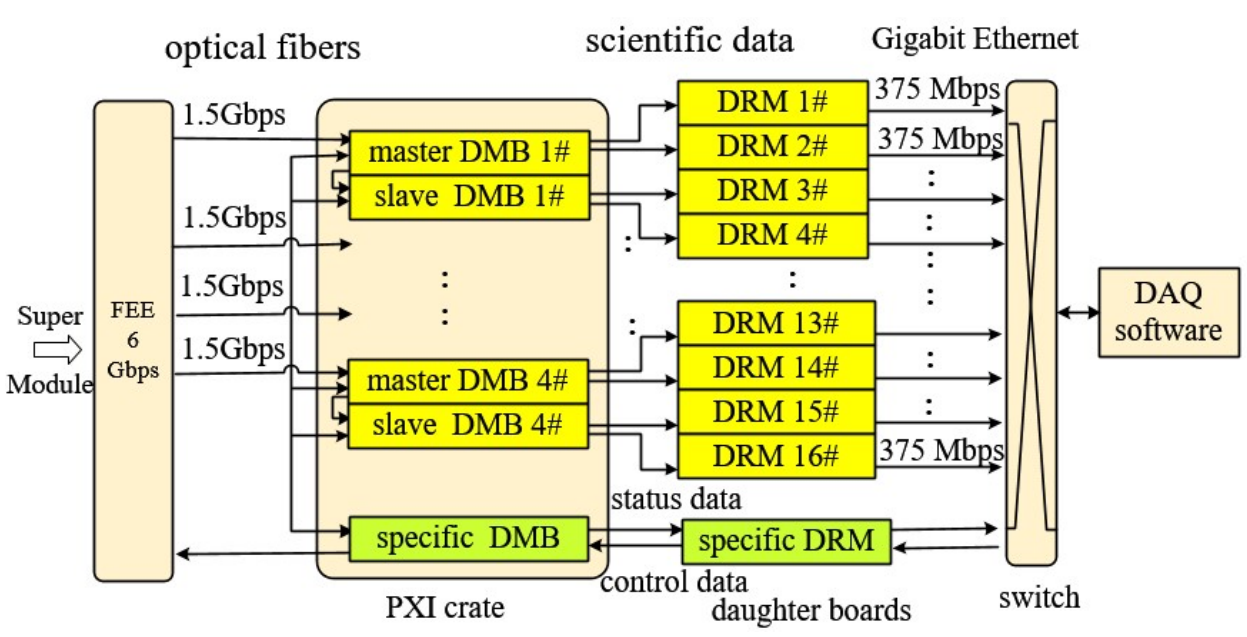

Fig. 3. The architecture of back-end readout electronics.

In [11], a test indicates that the transmission capacity of one DRM is about 550 Mbps, and it is limited by the single-core processing performance of the microprocessor in the SoC FPGA of the DRM [11].

There are three types of data: scientific data containing TDC time measurements, status data indicating whether the circuit modules work correctly, and control data containing commands for the readout electronics. The maximal rate of the scientific data is about 6 Gbps whereas the rate of the status data and the control data is too small and can be ignored. Therefore the critical problem is to transmit the scientific data received from STS at the rate of 6 Gbps.

The DMBs and DRMs are divided into four groups, each of which consists of two DMBs and four DRMs and is responsible for data transmission at the maximal rate of 1.5 Gbps. First, the scientific data and status data of the STS are mixed at the STS and then sent to four parallel master DMBs via optical fibers at the maximal rate of 1.5 Gbps per link. The optical fiber cuts off an electronic connection between the front-end electronics and back-end electronics, which makes a long-distance transmission available. Besides, the scientific data, as well as the status data of the STS, are separated at the master DMB and the status data are aggregated to the specific DMB (marked with green color in Fig. 3). Second, the master DMBs send half of the data to their slave DMBs. Next, each DMB, either master or slave, equally distributes the data to the two DRM daughter boards. After the first-grade equipartition between the master DMB and slave DMB, and the second-grade equipartition between two DRM daughter boards, all the scientific data are equally distributed to sixteen DRMs, each of which is expected to transmit the data at the maximal rate of 375 Mbps, which is within its transmission capacity of 550 Mbps. Finally, each DRM sends the data to the DAQ software through the standard TCP/IP Gigabit Ethernet.

The specific DMB and DRM, which have the same hardware as the other DMBs and DRMs, are responsible for the status and control data transmission. The status data are aggregated into the specific DMB via the crate backplane and then uploaded to the DAQ software by the specific DRM. As for the control data, they are firstly downloaded from the DAQ software to the specific DMB. Next, they are fanned out to other DMBs via the crate backplane and then sent to STS via optical fiber finally.

## III. Hardware Design Of DMB And DRM

The DMB and DRM denote the hardware support for the readout electronics. The structure of the DMB and DRM is shown in Fig. 4, and the corresponding photo is shown in Fig. 5.

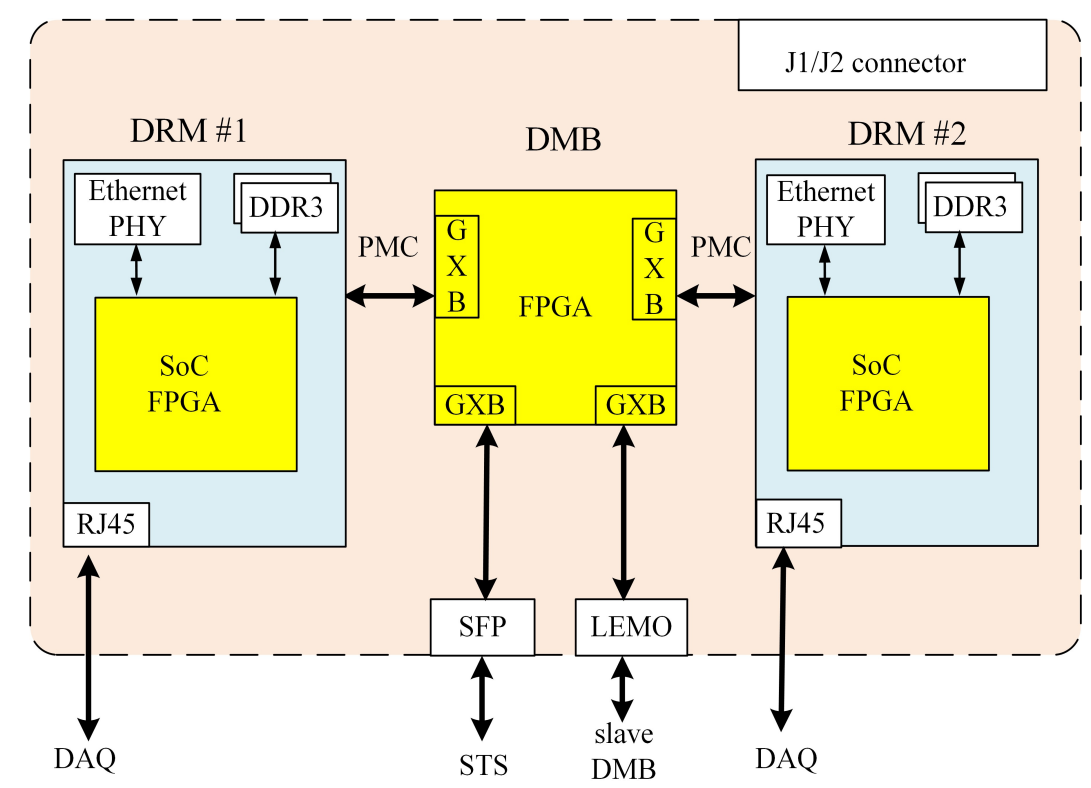

Fig. 4. Structure of DMB and DRM.

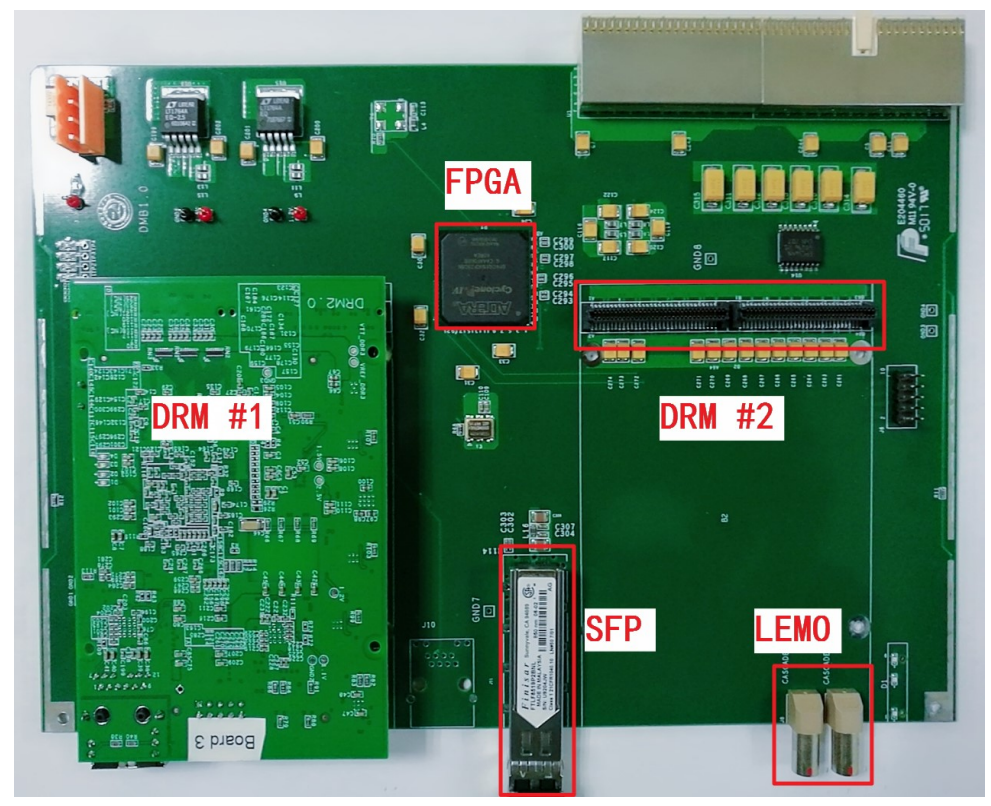

Fig. 5. Photo of DMB and DRM

### A. Hardware design of DMB

The DMB is designed as a standard 6U PXI plugin. A Cyclone IV GX series FPGA (EP4CGX30CF23C8) with four integrated four Gigabit transceiver blocks (GXBs) is utilized as the readout controller [12]. The GXB can support a high-speed and reliable data series transmission at the maximal data rate of 3.125 Gbps. As shown in Fig. 4, one of the GXBs of the master DMB is used to receive data from the STS through the optical fiber. The SFP transceiver is adopted for the photoelectric conversion between the STS and the master DMB. Another GXB is used to distribute half of the data from the master DMB to the slave DMB. A pair of differential LEMO connectors is adopted as the physical connector between the master DMB and slave DMB, which are exceptionally reliable and robust owing to the "Push-Pull" self-latching system [13]. The last two GXBs are used to transmit data to the two DRM daughter boards. The PMC connector is adopted as the interface connector between the DMB and DRM.

### B. Hardware design of DRM

The DRM is designed to support the Gigabit Ethernet transmission. Data from the DMB are received by the FPGA of the DRM through the GXB and then sent to the DAQ software through the standard TCP/IP Gigabit Ethernet. The DRM mainly comprises a SoC FPGA chip, a DDR3 RAM, and an Ethernet module. A Cyclone V series FPGA (5CSXFC2C6U23C8N) is utilized as the readout controller [14]; it is implemented by the SoC FPGA technique which packages the FPGA and ARM processor in a single die. Also, it has the advantages of higher integration, smaller board size, and higher interaction bandwidth between the ARM processor and FPGA. The embedded real-time Linux software, which runs on a hard processor system, packages the data received by the FPGA into the Ethernet format data with the TCP/IP protocol, and then sends the data to the DAQ software. The DAQ software is developed under Linux operating system running on the back-end computer. The DAQ software is responsible for data collection, event building, status monitoring, and system control [11].

## IV. Preliminary Test

A system level test was performed to verify the performance of the proposed readout electronics method. Following the explanation in Section II, in the test, the STS data were first divided into four subflows and then transmitted to four parallel master DMBs through the optical fibers. Since the readout architecture of the four subflows was the same (as shown in Fig. 3), the test system was simplified, containing only one subflow. The test system block diagram is shown in Fig. 6. The expected transmission capacity of the test system was 1.5 Gbps. The test data were generated by the FPGA in the STS, and the data rate was about 1.6 Gbps, which was a little higher than the expected maximal data rate for the necessary margin. The DAQ software recorded the rate of the received data every 10 s during the whole test time of four hours. The test result is presented in Fig. 7. As Fig. 7 shows, the test system succeeded in transmitting data at the rate of about 1.6 Gbps, which indicated that the proposed readout electronics met the requirement for transmitting data at the maximal rate of 6 Gbps.

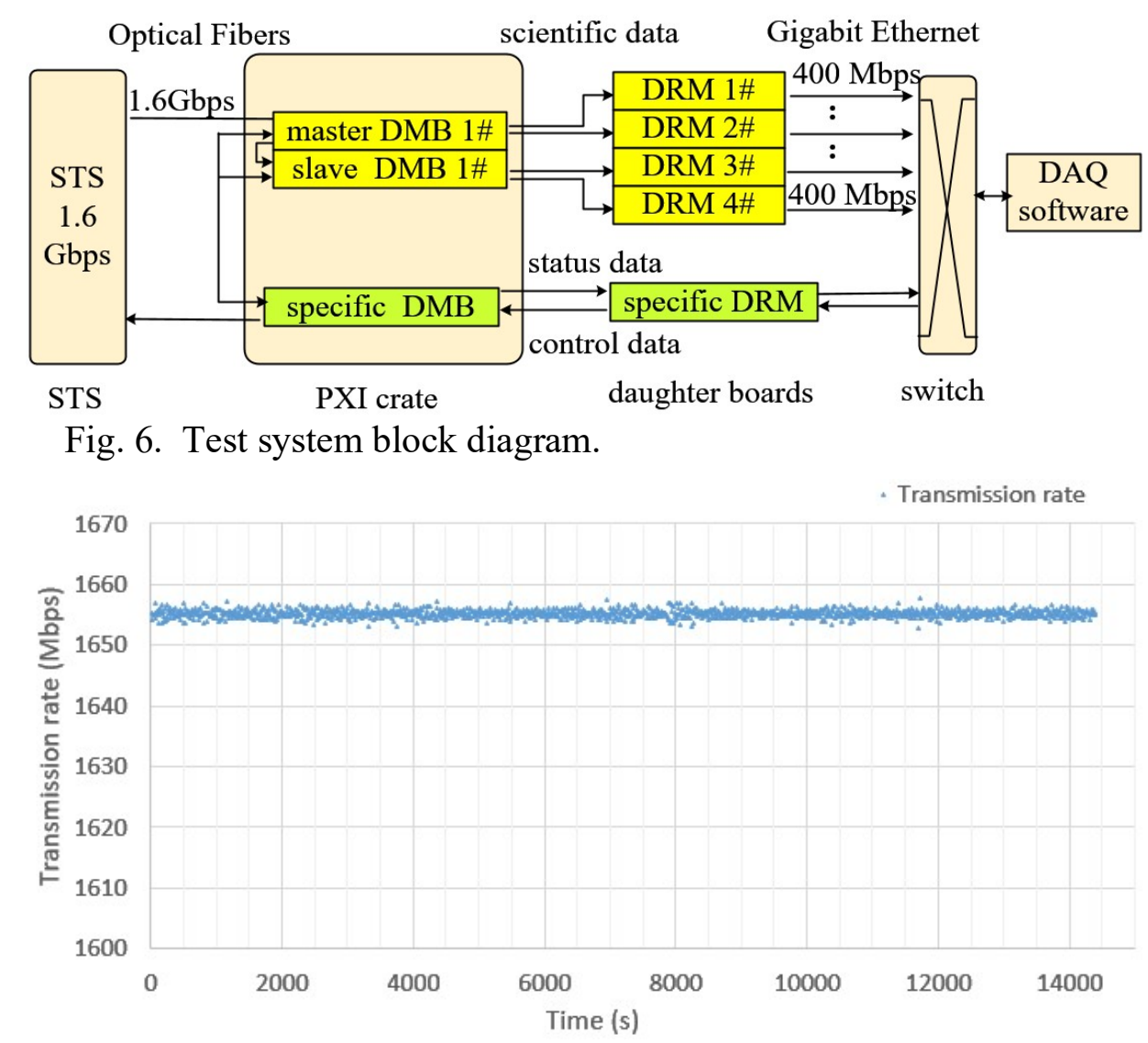

Fig. 6. Test system block diagram.

Fig. 7. Test result of the data transmission rate.

## V. Conclusion

A super module for the CBM-TOF experiment requires a high speed, high density, and high precision, bringing a tremendous challenge to the readout electronics for transmitting data at the maximal speed of 6 Gbps. Due to the limitation of the crate backplane bandwidth, the traditional VME or PXI technique cannot meet the requirement for high-speed data transmission. This paper proposes a new readout method utilizing a parallel architecture based on the Gigabit Ethernet. Using the proposed readout method, the data from the front-end electronics are sent to the readout module groups consisted of the DMBs and DRMs. The DMBs equally distribute data to sixteen DRMs step by step so that the data throughput of a single DRM is within its transmission capacity. The DRMs transmit data to the DAQ software through standard Gigabit Ethernet at the maximal rate of 375 Mbps. The preliminary test shows that one of the four parallel DMB-DRM groups can transmit data at the speed of 1.6 Gbps, which indicates the overall readout system can transmit data at the maximal speed of 6 Gbps. Therefore, the proposed readout method can meet the requirement for quality evaluation of a supper module in the CBM-TOF.

Furthermore, the proposed readout method has the advantage of good scalability. Thus, besides the CBM-TOF super module quality evaluation, this readout method can be applied to other related physical experiments having a variable data rate.